\input{aipcheck}

\documentclass[
    ,final            
  ]
  {aipproc}

\layoutstyle{8x11double}

\begin{document}

\title{Setup for shot noise measurements in carbon nanotubes}


\author{Fan Wu}{
   address={Low Temperature Laboratory, Helsinki University of Technology, P.O. Box 2200, 02015 TKK, Finland}
}

\author{Leif Roschier}{
   address={Low Temperature Laboratory, Helsinki University of Technology, P.O. Box 2200, 02015 TKK, Finland}
}
\author{Taku Tsuneta}{
   address={Low Temperature Laboratory, Helsinki University of Technology, P.O. Box 2200, 02015 TKK, Finland}
}

\author{Mikko Paalanen}{
   address={Low Temperature Laboratory, Helsinki University of Technology, P.O. Box 2200, 02015 TKK, Finland}
}

\author{Taihong Wang}{
  address={Institute of Physics, Chinese Academy of Sciences, Beijing 100080, China}
}
\author{Pertti Hakonen}{
   address={Low Temperature Laboratory, Helsinki University of Technology, P.O. Box 2200, 02015 TKK, Finland}
}

\begin{abstract}
We have constructed a noise measurement setup for high impedance
carbon nanotube samples. Our setup, working in the frequency range
of 600 - 900 MHz, takes advantage of the fact that the shot noise
power is reasonably large for high impedance sources so that
relatively large, fixed non-matching conditions can be tolerated.
\end{abstract}

\maketitle


Noise measurements in out-of-equilibrium conditions can be
employed to acquire extra information on top of that obtained from
ordinary conductance measurements, the data of which are related
to equilibrium noise by the fluctuation dissipation theorem. Such
noise measurements, however, are hindered by the ubiquitous 1/f
noise, in respect to which
 carbon nanotube devices make no
exception$\left[ 1-9 \right]$. Collins {\it et al}.\ measured 1/$f$
noise of several samples of single-walled nanotubes (SWNT), and
found them to be so noisy that their use as electronic components is
compromised\cite{Collins}, at least at room temperature. On the
contrary, extremely good low-frequency noise properties have been
achieved for single electron transistors made out of multiwalled
carbon nanotubes (MWNT) by Roschier {\it et al.}\cite{free}.

Altogether, on the basis of the noise measurements performed so far,
it has become apparent that standard low-noise experimental
techniques, reaching only up to 10 kHz due to $RC$ cut-off problems,
are not sufficient to study shot noise phenomena in carbon
nanotubes. One approach to circumvent the limitations due to RC
cut-off is to use resonant techniques to compensate the lead
capacitance. Alternatively, microwave techniques with impedance
matching may be employed. Both methods call for high-frequency, low
temperature amplifiers, working in either the MHz or even the GHz
regime. Our new noise measurement setup employs microwave techniques
but differs slightly from the standard solution by allowing for
unmatching conditions for the sample.

The heart of our setup is a cooled, home-made HEMT preamplifier
that operates in the frequency range of range of 600 - 950
MHz\cite{LeifCryo04}. The average noise temperature of this
low-noise amplifier (LNA) over the measurement bandwidth of
600-900 MHz is about 4 K. Under strong unmatching conditions, the
back action noise of the preamplifier is fully reflected back to
the amplifier, independent of the sample impedance, and therefore
this gives only a constant shift in the level of integrated noise
at the output. The setup is illustrated in Fig. 1.

\begin{figure}

        \includegraphics[height=.2\textheight]{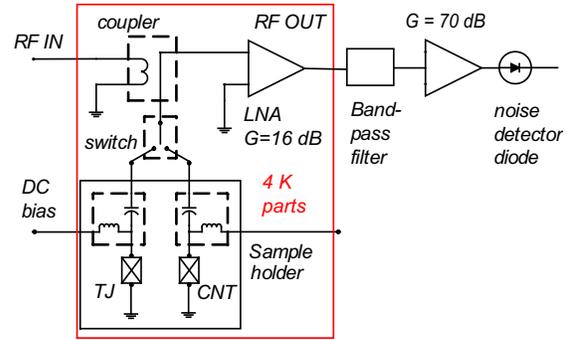}

    \caption{Principle of the shot noise measurement setup.
    A microwave frequency switch is
    employed to select either the studied object or a
    tunnel junction noise source. The  noise integrated over the bandwidth
    600 - 900 MHz is read either from the output of the Schottky
    diode detector or from the output of the lock-in amplifier if
    AC-modulation of current is applied. For further details, see
    text.
    } \label{setup}
\end{figure}

The main goal in our setup is to measure reliably the Fano-factor
$F$ which relates the measured noise $S_I$ to the full shot noise
$2e<I>$ of a Poissonian process: $F= S_I/(2e<I>)$, where $<I>$ is
the average current. This is a characteristic number for
mesoscopic samples in general \cite{BB}. The Fano-factor is
connected to quantum partition noise, which for a single transport
channel is given by

\begin{equation}\label{quantumnoise}
  <(\delta I)^{2}> =2e < I > (1-\tau),
\end{equation}
where transmission coefficient is denoted by $\tau$. In a
multichannel system, sum over the transmission channels is taken.
For example, for a mesoscopic diffusive wire, one obtains $F=1/3$
without interaction or hot electron effects. For a tunnel junction,
$F = 1$. In fact, by having a high impedance tunnel junction and a
microwave switch, this latter fact is employed to calibrate the
sensitivity of our noise setup.

In our setup, bias-tees are used to allow voltage biasing of the
sample and to measure the IV-characteristics while the noise
measurement at microwave frequencies is going on. The amplification
by the 4.2 K amplifier enhanced by three room-temperature amplifiers
in series in order to make the signal level sufficient for a
Schottky-diode power detector. Band width limitation is employed
before the detector in order to cut off the extra noise due to the
rather wide band width of the room temperature amplifiers (6 GHz for
MITEQ SMC03 and 20 GHz for SMC05). Small AC-modulation is typically
applied on top of the DC-bias so that drift in the noise level is
eliminated by using lock-in techniques after detection. The full
noise can then be obtained by numerical integration. This scheme,
however, does not eliminate the drift in the gain of the amplifiers.
Temperature stabilization of the LNAs, especially the room
temperature ones, has been found to decrease this drift
substantially.

The noise power of a source having an impedance $R$ is given by
\begin{equation}\label{Pnoise}
  P_{noise}=S_I R=4k_{B}T(1-F)+F2eIR \coth \frac{eV}{2k_{B}T} .
\end{equation}
This formula shows that the thermal noise and the shot noise are
intermixed in a manner depending on the Fano-factor, which makes a
separation of the contributions of thermal and shot noise
complicated. Since the preamplifier is matched to a transmission
line having the impedance of $Z_0 =50$ $\Omega$, the coupling from
the source to the preamplifier is governed by the mismatch between
the transmission line and the source. This is governed by reflection
coefficient $\Gamma=(R-Z_0)/(R+Z_0)$ in which $R$ denotes the
small-signal resistance of the sample at the operating point. Thus,
the noise power $P_{measure}$ coupled to the preamplifier becomes

\begin{equation}\label{Pmeasure}
  P_{measure}=P_{noise}(1-|\Gamma|^{2})\sim F 8eI
  Z_{0},
\end{equation}
where the latter form is valid in the regime $FeV>>k_B T$.
Therefore, for samples with only weakly non-linear IV-curves, the
Fano factor can be obtained directly as the ratio of the slopes of
the noise vs. current curves measured for the tunnel junction and
the nanotube samples.

Fig. 2 illustrates the resolution achieved on a tunnel junction
sample of resistance $R_T = 7.7$ k$\Omega$. The signal to noise
ratio is clearly so good that the main limitation of our method
comes from the requirement of $FeV>>k_B T$. As worked out by Spietz
et al \cite{spietz04}, the cross-over regime between thermal and
shot noise can be employed to determine absolute temperature of the
sample, and by using the fitted lines we get $T=4.2$ K as expected.

\begin{figure}

        \includegraphics[height=.25\textheight]{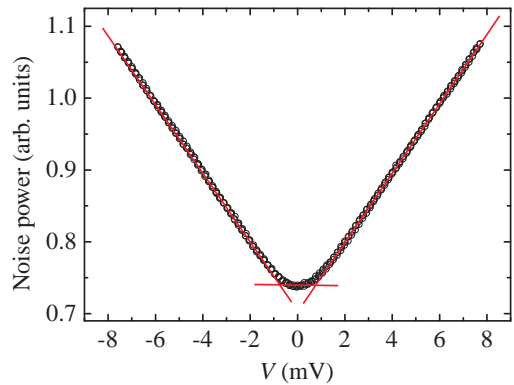}

    \caption{Noise vs. bias current measured on a tunnel junction with $R_T=7.7$
    k$\Omega$. The tunnel junction was made of Al/AlO$_x$/Al
    using standard two-angle shadow evaporation.
    } \label{setup}
\end{figure}


\begin{theacknowledgments}
It is a pleasure to thank T. Lehtinen, L. Korhonen, M.
Sillanp\"a\"a, and R. Tarkiainen for valuable help. This work was
supported by Academy of Finland, and by the Vaisala foundation.
\end{theacknowledgments}



\end{document}